\documentclass[aps,prl,showkeys,floatfix,amsmath,superscriptaddress, amssymb,reprint]{revtex4-1}
\usepackage{graphicx} 
\usepackage{bm}
\usepackage{color,mathtools,bbm} 
\usepackage{wasysym}    
  \usepackage{lineno}
\usepackage{longtable}
\usepackage{footnotehyper}
\makesavenoteenv{longtable}
\usepackage{xcolor}
\usepackage{soul}
\usepackage{array}
\usepackage[normalem]{ulem}
\usepackage{mathptmx}
\usepackage{etoolbox}
\usepackage{todonotes}
\usepackage{hypernat}
\usepackage{nameref}
\usepackage{wrapfig} 
\usepackage{adjustbox}
\usepackage{multirow}
\usepackage{braket}
      
\usepackage{hyperref}
\begin{document}

\title{Intrinsic Chern Half-Metal with High Anomalous Hall Conductivity in 2D BaNiCl$_3$}

\author{Chidiebere I. Nwaogbo}
\affiliation{Department of Physics, Lehigh University, Bethlehem, PA 18015, USA}
\author{Chinedu E. Ekuma}
\altaffiliation{che218@lehigh.edu}
\affiliation{Department of Physics, Lehigh University, Bethlehem, PA 18015, USA}

\date{\today}

\begin{abstract}
Two-dimensional (2D) half-metals offer complete spin polarization at the Fermi level, making them candidates for dissipationless spin transport. Yet intrinsic 2D half-metals exhibiting robust topological features, particularly large Chern number anomalous Hall conductivities, remain exceptionally rare. Using first-principles calculations, we identify atomically thin BaNiCl$_3$, a layered halide perovskite (perovskene), as a topological half-metal. It exhibits a high Chern number ($C \ge 2$), a large anomalous Hall conductivity of 316~$\Omega^{-1},\mathrm{cm}^{-1}$, and a Fermi velocity of $\approx 0.78 \times 10^6$ m/s. The coexistence of complete spin polarization and high carrier velocity suggests low-dissipation spin transport. Spin-orbit coupling opens a sizable topological gap of $\sim 20\,$ meV, yielding nontrivial Berry curvature and enhancing the anomalous Hall response. Ferromagnetism is stabilized by the Ni$^{2+}$ ($d^8$) configuration and Cl-mediated superexchange, supporting magnetic ordering at elevated temperatures. These results establish BaNiCl$_3$ as a rare intrinsic Chern half-metal, with potential applications in quantum and spintronic technologies. 
\end{abstract}
  
\maketitle

Topological phases with strong Berry curvature effects continue to attract interest for their potential in low-dissipation electronics and quantum information technologies~\cite{hasan2015topological,kane2005quantum,RevModPhys.89.040502}. Among these, materials with large anomalous Hall conductivity (AHC) are especially promising, as their intrinsic Berry curvature enables efficient transverse charge transport~\cite{PhysRevLett.61.2015,thouless1982quantized,li2023quantum}. Maximizing AHC requires tuning the electronic structure to enhance Berry curvature near the Fermi level—achievable through magnetic ordering to break time-reversal symmetry, spin-orbit coupling (SOC) to induce asymmetric dispersion, and symmetry engineering to generate momentum-space singularities~\cite{wolf2001spintronics,vzutic2004spintronics,hu2015chern,slager2013space}. In particular, phases with high Chern number ($C \ge 2$) exhibit stronger Berry curvature accumulation due to contributions from multiple topological bands, often enabling large Hall responses even in metallic systems. Representative systems include kagome lattices, magnetic Weyl semimetals, and transition-metal compounds with nontrivial band topology~\cite{roychowdhury2024enhancement,wang2024topological,liu2019magnetic}. While such materials offer a promising path toward energy-efficient spintronics, their implementation is often hindered by low Curie temperatures or competing electronic states that suppress AHC. Concurrently, half-metals (HMs), defined by complete spin polarization at the Fermi level, have emerged as key candidates for spin injection and filtering in spin-based logic~\cite{ishizuka2012dirac,park1998direct,de1983new,wu2015atomically,he2016unusual}. Although half-metallicity has been realized in 3D Heusler compounds and perovskites, achieving robust spin-polarized gaps and high $T_C$ in truly two-dimensional (2D) systems remains challenging~\cite{xiang2006one,singh2015theoretical,zhou2011magnetism,liu2022realization}.

Recently, combining half-metallicity with nontrivial topology has emerged as a strategy for realizing fully spin-polarized Chern states, termed Chern half-metals~\cite{hu2015chern,zhang2022proximity,wang2023topological}. These systems feature gapless bands in one spin channel and topologically nontrivial gaps in the other, enabling spin-polarized and dissipationless conduction. However, intrinsic 2D Chern half-metals remain exceedingly rare due to the delicate balance of exchange splitting, SOC, and lattice symmetry. Layered perovskites provide a tunable platform where orbital hybridization, transition-metal identity, and ligand coordination can be engineered to stabilize such phases~\cite{kato2002metallic,chen2019direct}. While several perovskite-based half-metals, such as manganites and Co-based oxides, exhibit structural flexibility and elevated $T_C$\cite{de1983new,wurmehl2005geometric,govind2020structural,zhang2012half}, realizing a topological gap concurrent with half-metallicity remains difficult, as strong magnetism and wide gaps often compete~\cite{deng2020molecular,liu2020quadruple}.

\begin{figure}[htb!]
	\centering
	\includegraphics[trim = 0mm 0mm 0mm 0mm,width=\linewidth,clip=true]{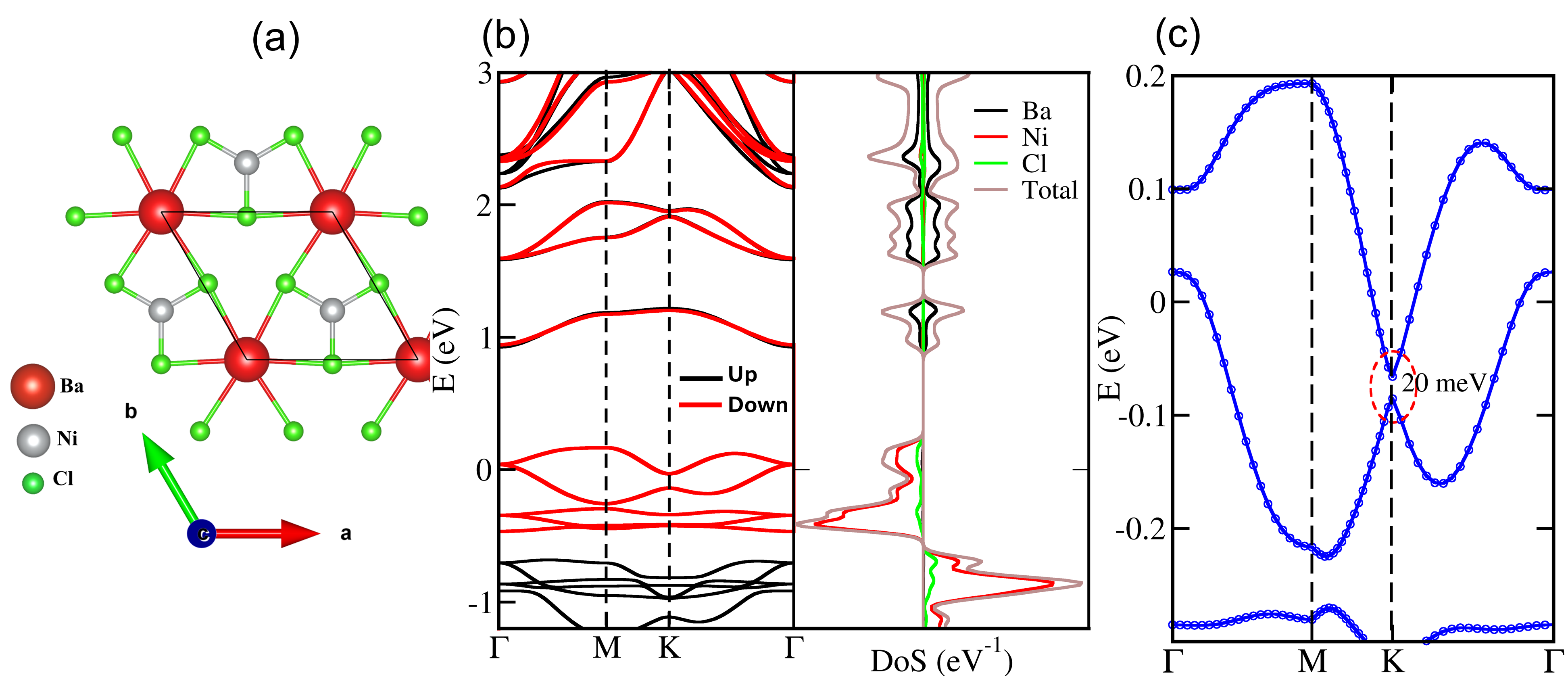} 
	\caption{ (a) Crystal structure of monolayer BaNiCl$_3$ showing the unit cell. (b) Spin-polarized electronic band structure and projected density of states (PDOS) without spin-orbit coupling (SOC), showing a half-metallic ground state where one spin channel is insulating while the other remains metallic.(c) Electronic band structure with spin-orbit coupling (SOC), highlighting the SOC-induced topological bandgap of $\approx20\,$meV at the \textit{K} high-symmetry point. This bandgap signifies the transition to a topological phase with a Chern number \(C = 2\), as evidenced by the emergence of chiral edge states.} 
	\label{fig:Figure1}
\end{figure}

Although prior studies have explored Chern phases via external modifications—such as magnetic doping, adatom decoration in Co/graphene systems~\cite{hu2015chern}, heterostructuring, or strong spin–orbit coupling—the realization of a material that intrinsically hosts a high Chern number ($C \ge 2$) and half-metallic edge states remains rare. Here, we demonstrate from first-principles calculations that monolayer BaNiCl$_3$ addresses key challenges in realizing intrinsic Chern half-metals. While the bulk compound exhibits structural tunability and strong crystal-field effects, exfoliation to a monolayer leads to a reorganization of the electronic structure, stabilizing an intrinsic half-metallic state with topologically nontrivial bands. The Ni$^{2+}$ ($d^8$) configuration drives an interplay among crystal field, electronic correlations, and spin–orbit coupling, yielding a spin-polarized band structure where one spin channel remains metallic while the other becomes insulating. Inclusion of SOC opens a topological bandgap of $\sim 20$ meV at the $\mathbf{K}$ point. Integration of the Berry curvature over the Brillouin zone—validated by discontinuities in the Wannier charge centers and the emergence of chiral edge states—confirms a Chern number of $C = 2$. Additionally, the system exhibits a high Fermi velocity of $\sim 0.78 \times 10^6$ m/s, supporting efficient spin-polarized transport. The combination of intrinsic half-metallicity, robust topology, SOC-induced gaps, and high-mobility carriers positions monolayer BaNiCl$_3$ as a promising candidate for topological spintronic applications.

To investigate these properties, we performed first-principles calculations~\cite{kohn1996density} combined with a Wannier-derived low-energy Hamiltonian to capture the topological properties. Monolayer BaNiCl$3$ was selected from the \textsc{perovskene} database~\cite{perovskene-github,perovskene-pub}. All structural relaxations and electronic structure calculations were carried out using the Vienna \emph{ab initio} Simulation Package~\cite{kresse1996efficiency}, with the projector augmented-wave method and the Perdew–Burke–Ernzerhof exchange–correlation functional~\cite{perdew1996generalized}. A plane-wave cutoff of 500 eV and a $\Gamma$-centered $5 \times 5 \times 1$ $\Gamma$-centered $k$-mesh were employed to sample the 2D Brillouin zone. A vacuum spacing of $\sim$20~\AA\ was introduced to prevent interlayer interactions. Total energies were converged to $10^{-7}$ eV, and forces were minimized below $0.02$ eV/\AA\ (see Figure \ref{fig:Figure1}a). Energetic (formation and exfoliation energy) and mechanical (Born stability) analyses are provided in the Supplemental Material (SM)\cite{supp}. Spin-polarized calculations were performed both with and without spin–orbit coupling. A low-energy Hamiltonian was constructed by projecting onto Ni $d$ and Cl $p$ orbitals within an energy window of $\pm 6.0$ eV around the Fermi level, using maximally localized Wannier functions~\cite{marzari2012maximally}. Topological properties, including the Berry curvature, Chern number, and edge spectra, were computed using \textsc{WannierTools}~\cite{wu2018wanniertools}. The Chern number was evaluated via $C = \frac{1}{2\pi}\int{\mathrm{BZ}} \Omega(\mathbf{k}) d^2\mathbf{k}$, where $\Omega(\mathbf{k})$ is the Berry curvature summed over occupied bands. Integration was performed over a dense $k$-mesh to accurately capture topological features, including discontinuities in the Wannier charge centers.

\begin{figure}[htb!]
	\centering
	\includegraphics[trim = 0mm 0mm 0mm 0mm,width=\linewidth,clip=true]{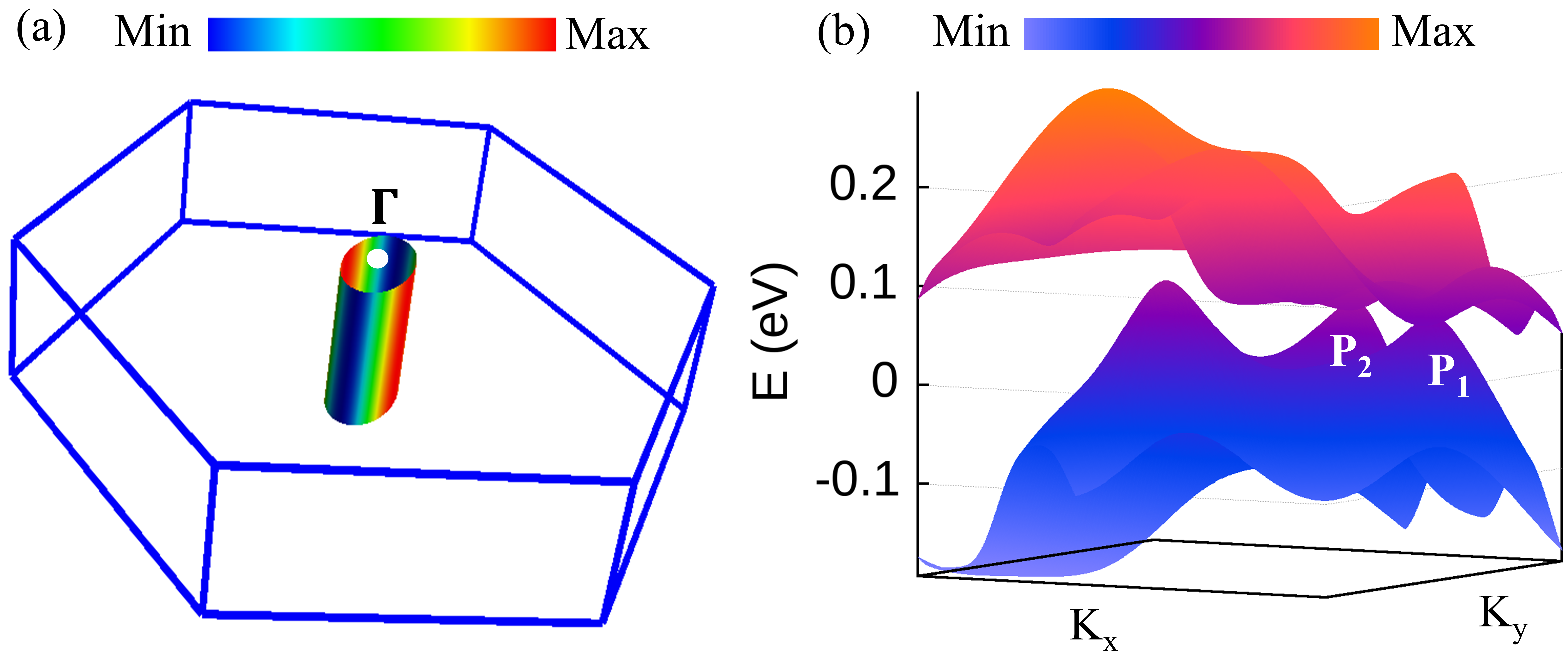} 
	\caption{(a) Fermi surface of 2D BaNiCl$_3$, centered around $\Gamma$ in the first Brillouin zone. The color scale represents the dispersion of the Fermi velocity. (b) Projected 3D representation of the band structure of 2D BaNiCl$_3$, plotted as a function of the in-plane momentum components $K_x$ and $K_y$. The color gradient represents the energy variation, with high-energy states in red and low-energy states in blue. The points labeled $P_1$ and $P_2$ indicate regions of linear dispersion, which host the topological features.} 
	\label{fig:fig2}
\end{figure}


The electronic band structure and density of states (DoS) for BaNiCl$_3$ are presented in Figure~\ref{fig:Figure1}(b-c). The band structure reveals a fully gapped state at the Fermi level in the spin-up channel, while the spin-down channel remains metallic, confirming its half-metallic nature. This arises from the strong hybridization between Ni-\textit{d} orbitals and Cl-\textit{p} orbitals, as observed in the projected band structure (Figure~S2)\cite{supp} and DoS (Figure~\ref{fig:Figure1}(b)), where the spin-down states are primarily composed of Ni-\textit{d} orbitals. The calculated PBE bandgap of 1.61 eV in the spin-up channel further emphasizes the half-metallicity, which increases to 2.25 eV upon incorporating an on-site Coulomb interaction (\textit{U} = 6.0 eV) using DFT+\textit{U}\cite{anisimov1997first,himmetoglu2014hubbard}. Spin-orbit coupling (SOC) induces significant modifications to the electronic structure opening a 20 meV direct bandgap at the $K$ point (Figure~\ref{fig:Figure1}(c)). In particular, the sharp slope of the nearly linear dispersion of the edge states suggest an exceptionally high Fermi velocity, evaluated as $\hbar v_F \approx \partial E(k)/\partial k$, yielding $v_F \approx 0.78 \times 10^6$ m/s. This value is comparable to that of graphene ($\sim 10^6$ m/s), highlighting ultrafast carrier dynamics in BaNiCl$_3$. Such a high Fermi velocity improves charge transport efficiency, reinforcing the material’s potential for high-speed, low-dissipation spintronics applications. The Fermi surface (Figure~\ref{fig:fig2}a) exhibits a cylindrical geometry concentrated near the center of the 2D Brillouin zone ($\Gamma$ point), indicative of strong in-plane dispersion with minimal out-of-plane interaction, which is consistent with its 2D electronic behavior. Figure~\ref{fig:fig2}(b) further captures the Dirac-like dispersion of the emergent topological states near the high-symmetry \(\mathbf{K}\) point, supporting the presence of highly mobile carriers and robust topologically nontrivial electronic states. The resulting band structure in the topological channel contains two linear dispersion regions, labeled $P_1$ and $P_2$, each exhibiting robust Berry curvature due to preserved crystal symmetries. Each gapped cone contributes a Chern number of $C = 1$, and their combined contributions yield a net Chern number of $C = 2$. The underlying lattice symmetry ensures that these multiple band inversions are protected in the topological channel while also distributing the Berry curvature evenly in momentum space, reinforcing the stability of the topological phase.

\begin{figure}[htb!]
	\centering
	\includegraphics[trim = 0mm 0mm 0mm 0mm,width=\linewidth,clip=true]{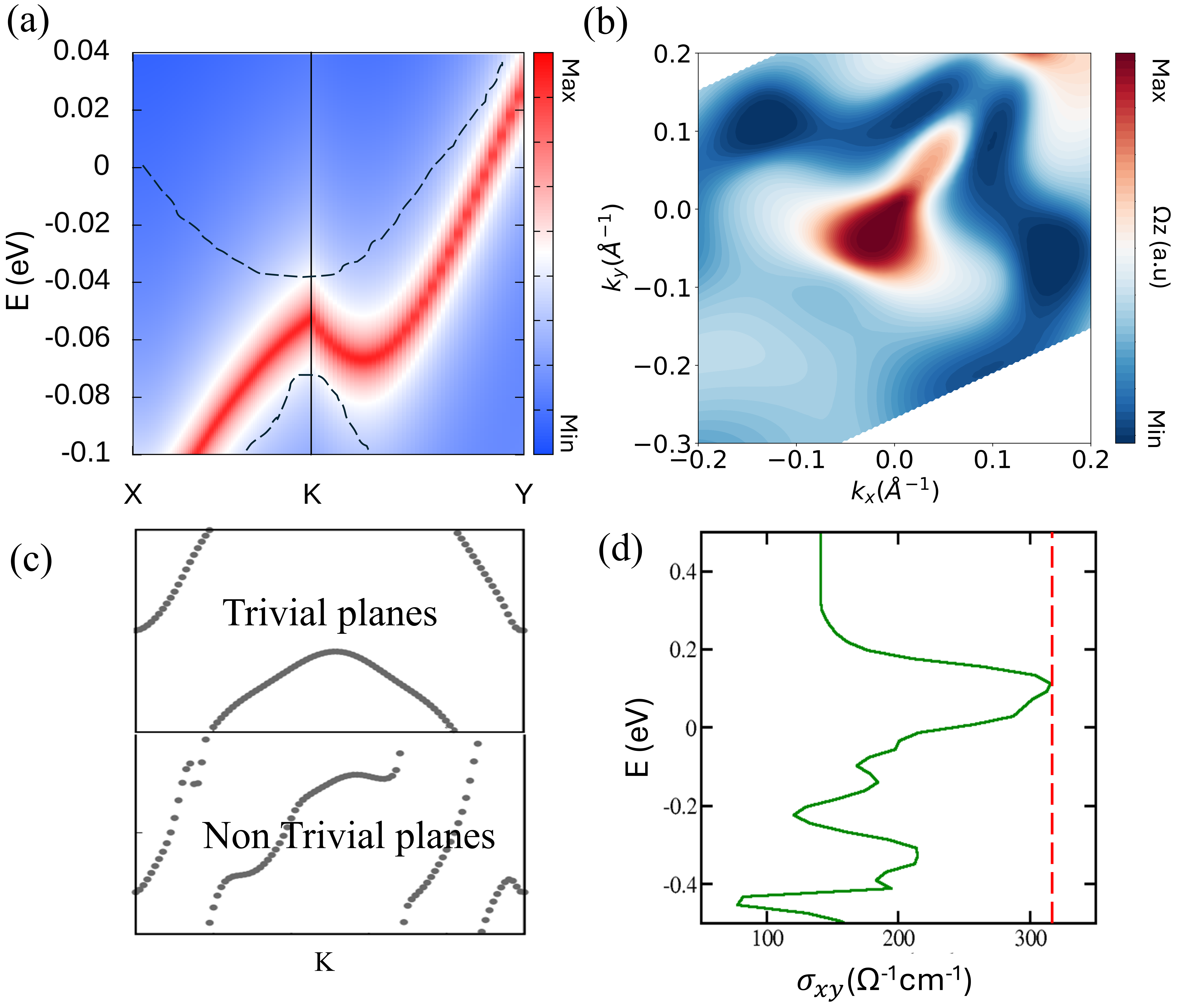} 
	\caption{(a) Topologically protected edge states, with dashed lines indicating a sketch of the bulk dispersion at the $K$ point. (b) Corresponding Berry curvature hotspot. (c) Evolution of the Wannier charge centers across the first Brillouin zone for trivial planes with $C = 0$ and nontrivial planes with ($C = 2$) (d) Intrinsic anomalous Hall conductivity (AHC), $\sigma_{xy}$ (in $\Omega^{-1} \text{cm}^{-1}$), as a function of energy for 2D BaNiCl$_3$. The red dashed line indicates the peak of the AHC.} 
	\label{fig:Figure3}
\end{figure}

\begin{figure}[b!]
	\centering
	\includegraphics[trim = 0mm 0mm 0mm 0mm,width=\linewidth,clip=true]{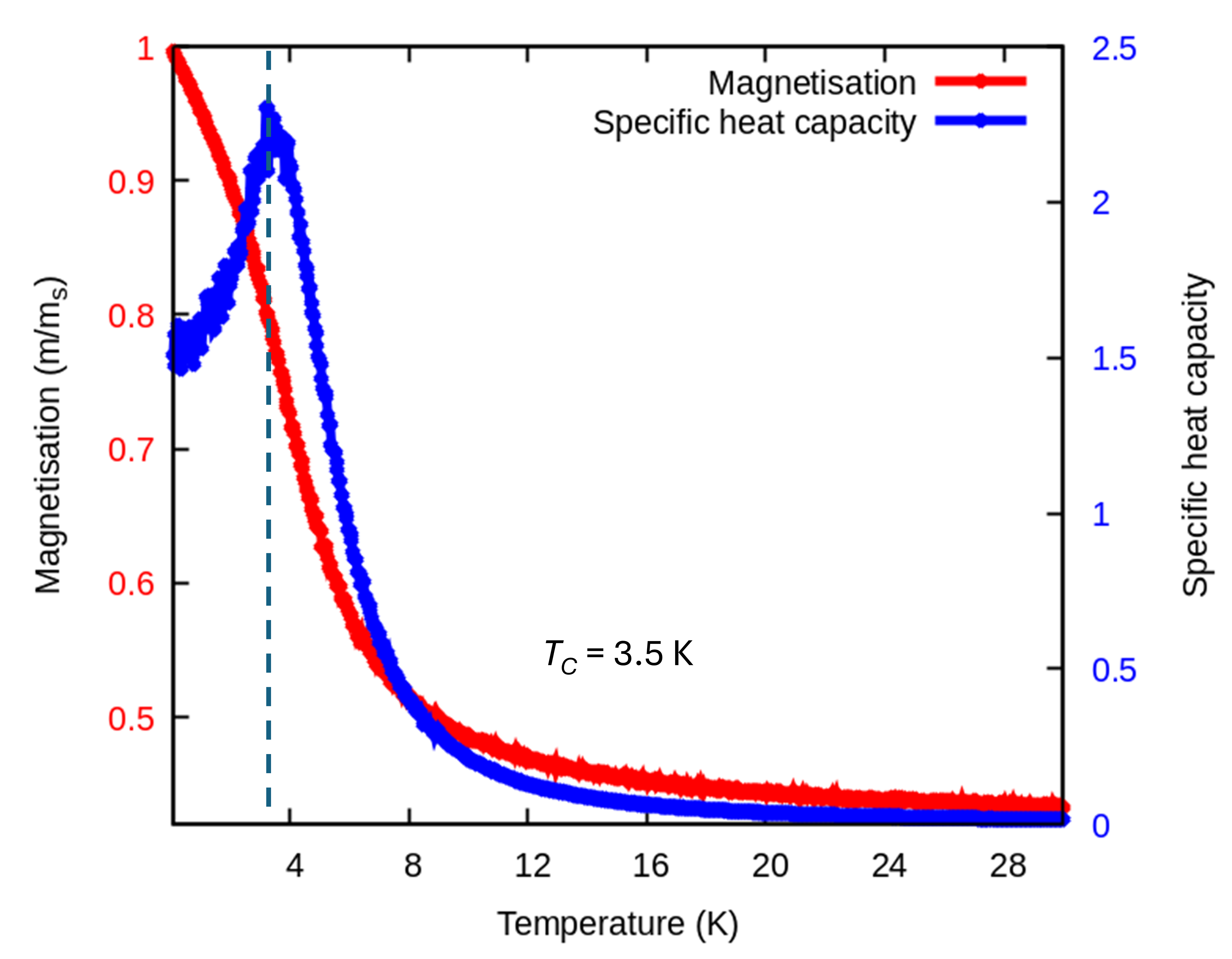} 
	\caption{Temperature dependence of the specific heat (blue) and normalized magnetization (red) for 2D BaNiCl$_3$.} 
	\label{fig4}
\end{figure}

Magnetic calculations reveal a total magnetic moment of 3.93,$\mu_B$ per unit cell in monolayer BaNiCl$3$. The ferromagnetic (FM) configuration constitutes the magnetic ground state, with an energy difference of $\Delta E = E{\mathrm{FM}} - E_{\mathrm{AFM}} \approx -90$ meV relative to the antiferromagnetic (AFM) state. The thermal stability of the magnetic ground state was assessed by estimating the Curie temperature $T_C$ via Monte Carlo simulations using the VAMPIRE code~\cite{evans2014atomistic}, based on the anisotropic 2D Heisenberg model: $H = -\frac{1}{2} J_1 \sum_{\langle i,j \rangle} \mathbf{S}_i \cdot \mathbf{S}_j - \frac{1}{2} J_2 \sum_{\langle\langle i,j \rangle\rangle} \mathbf{S}_i \cdot \mathbf{S}_j - K \sum_i |S_i^z|^2$, where $J_1$ and $J_2$ are the nearest- and next-nearest-neighbor exchange interactions, $\mathbf{S}_i$ is the spin vector, and $K$ is the single-ion anisotropy. The exchange parameters were extracted from total energy differences among FM, G-type AFM, and A-type AFM configurations (see Table S1~\cite{supp}). Positive values of $J$ confirm ferromagnetic coupling, although their small magnitudes indicate weak exchange interactions and a correspondingly low $T_C$. Temperature-dependent magnetic susceptibility $\chi(T)$, specific heat, and normalized magnetization $M(T)/M_0$ (Figure~\ref{fig4} and Figure~S3\cite{supp}) exhibit clear signatures of a magnetic phase transition, with Monte Carlo simulations yielding $T_C \approx 3.5$ K and a complementary mean-field estimate of $T_C \approx 11$ K. Although $T_C$ lies well below room temperature, it may be enhanced through strain engineering or interfacial exchange coupling with ferromagnetic substrates. The magnetocrystalline anisotropy energy, $\mathrm{MAE} = E_{\parallel} - E_{\perp}$, is calculated to be 5 meV per formula unit, consistent with other 2D half-metals~\cite{hu2015chern,wang2023exploitable,bhardwaj2023magnetocrystalline}. The positive MAE favors out-of-plane magnetization, breaking time-reversal symmetry below $T_C$ and enabling symmetry-protected topological phases. Its magnitude also implies that substantial external fields are required to reorient the magnetic axis, indicating robust anisotropy. 

To confirm the topological nature of 2D BaNiCl$_3$, we analyzed the Wannier charge center (WCCs) evolution (Figures~\ref{fig:Figure3}(c). As shown in Figure~\ref{fig:Figure3}(c), the WCCs exhibit discontinuous jumps, confirming a nonzero Chern number \(C=2\) for the topological spin channel. In contrast, planes with trivial topology (with \(C=0\)) display smooth, continuous WCC evolution without net winding. The intrinsic anomalous Hall conductivity (AHC) of 2D BaNiCl$_3$ is plotted in Figures~\ref{fig:Figure3}(d), revealing a plateau near the Fermi level with $\sigma_{xy} \approx 316~\Omega^{-1}\text{cm}^{-1}$. This peak results from enhanced Berry curvature contributions due to SOC and ferromagnetic ordering. As shown in Fig.~\ref{fig:Figure3}(b), the dominant contribution to $\sigma_{xy}$ arises from the Weyl point at \textit{K}, where the Berry curvature exhibits a pronounced singularity. Previous studies have shown that 2D ferromagnetic Dirac materials can exhibit intrinsic $\sigma_{xy}$ values up to 300 $\Omega^{-1}\text{cm}^{-1}$ due to strong SOC and time-reversal symmetry (TRS) breaking \cite{yang2023tuning}. BaNiCl$_3$ demonstrates a remarkable AHC of 316~\(\Omega^{-1} \text{cm}^{-1}\), surpassing known 2D ferromagnetic Dirac materials such as FeCl$_2$ (300~\(\Omega^{-1} \text{cm}^{-1}\)) and exceeding several Heusler compounds, including Gd$_4$Sb$_3$ (233~\(\Omega^{-1} \text{cm}^{-1}\)) and Co$_2$TiSn (284~\(\Omega^{-1} \text{cm}^{-1}\))~\cite{han2024anomalous, roy2020anomalous}. Unlike bulk ferrimagnetic half-metals that achieve higher AHC values, such as Co$_{1/3}$NbS$_2$ (400~\(\Omega^{-1} \text{cm}^{-1}\)) and Cs$_2$Co$_3$S$_4$ (500~\(\Omega^{-1} \text{cm}^{-1}\))~\cite{tenasini2020giant, acharya2024large}, BaNiCl$_3$ achieves this exceptional performance within a purely 2D framework. Additionally, BaNiCl$_3$ offers a rare combination of intrinsic half-metallicity and a high Chern number ($C = 2$) without requiring external perturbations, positioning it as a superior candidate for integration into nanoscale spintronic devices where both high AHC and robust topological features are critical.

\begin{figure}[htb!]
	\centering
	\includegraphics[trim = 0mm 0mm 0mm 0mm,width=\linewidth,clip=true]{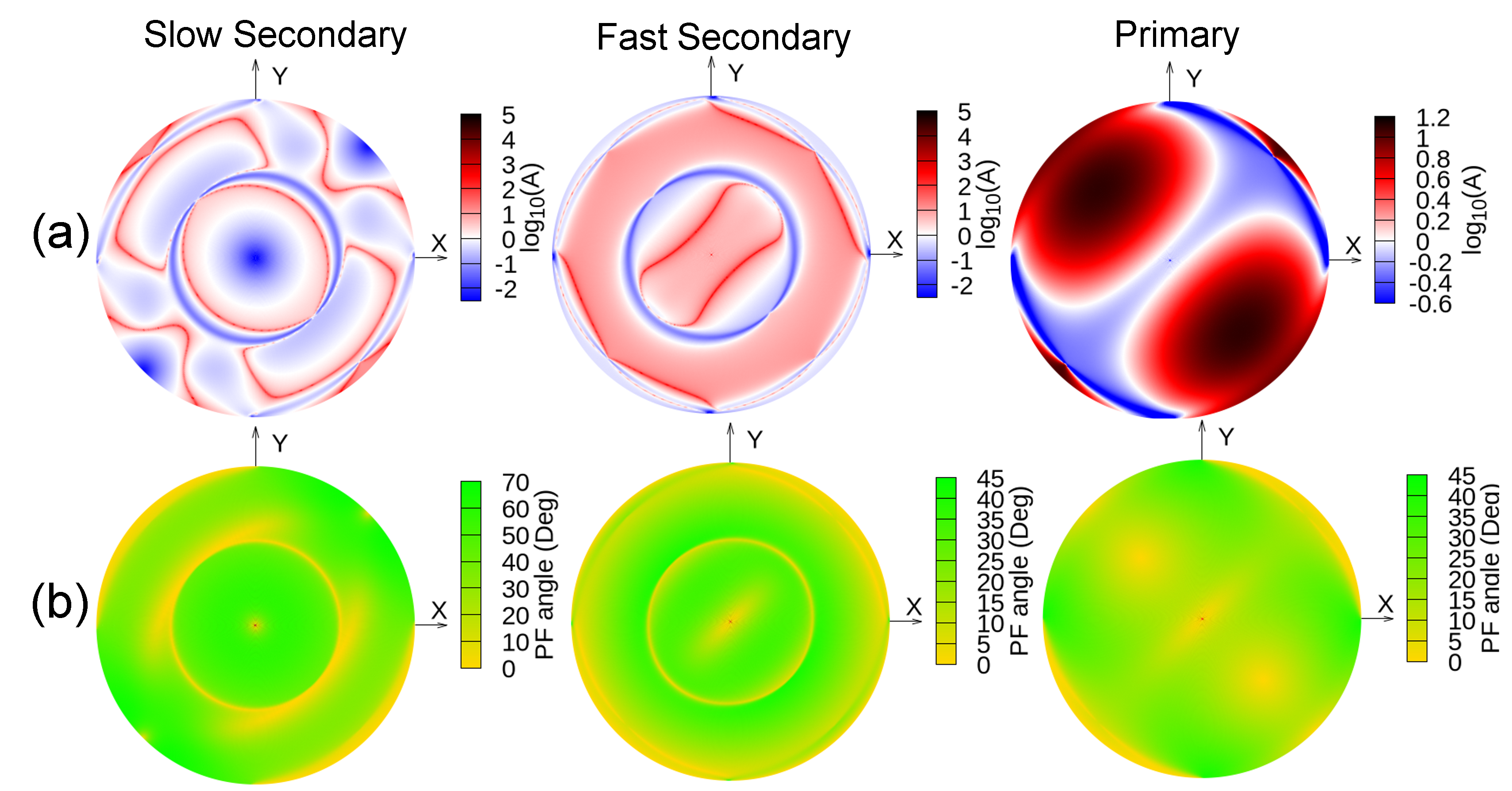} 
	\caption{Equal-area projection of (a) the enhancement factor and (b) the power flow angle (PF angle). In Figure~\ref{fig5}a, the observed directional anisotropic patterns in phonon energy transport, characterized by distinct ring-like lobes, correlate with regions of enhanced Berry curvature, suggesting the presence of topologically nontrivial phonon modes. In Figure~\ref{fig5}b, the PF angle shows deviations exceeding 70$^\circ$, which correlate with enhanced energy transport, reinforcing the presence of chiral-like phonon modes in 2D BaNiCl$_3$ Chern half-metal.} 
	\label{fig5}
\end{figure}

To further probe the topological character of monolayer BaNiCl$_3$, we examine its phononic properties, focusing on low-energy vibrational modes~\cite{liu2022elastool, EKUMA2024109161}. In the equal-area projection, the enhancement factor (Fig.\ref{fig5}(a)) provides a directional map of phonon transport anisotropy. The observed hemisphere-splitting patterns, ring-like high-intensity lobes, and suppressed low-intensity zones reflect pronounced anisotropy, likely originating from broken crystal symmetries and nontrivial band topology. These features suggest momentum-space regions of enhanced vibrational response that spatially align with electronic Berry curvature ``hot spots'' (see Figure~\ref{fig:Figure3}), indicating an interplay between topological electronic structure and lattice dynamics\cite{hassani2020topological}. In particular, we observe large angular deviations in the power flow (PF) of the slow secondary acoustic mode, with angles exceeding 70$^\circ$ (Figure~\ref{fig5}(b)). These deviations coincide with regions of high enhancement factors, indicating a directional concentration of phononic energy flow. This correlation may signal the emergence of chiral-like phonon propagation patterns influenced by the underlying electronic topology. While a rigorous topological classification of these modes requires explicit indicators such as phonon band inversions or edge-localized modes, the observed alignment of PF anomalies with Berry curvature regions supports the possibility of topology-driven phononic anisotropy in 2D BaNiCl$_3$.

In summary, monolayer BaNiCl$_3$ is identified as an intrinsic Chern half-metal that combines robust ferromagnetism, nontrivial band topology, and significant spin–orbit coupling. First-principles calculations reveal a half-metallic ground state, with SOC opening a topological gap at the $\mathbf{K}$ point and yielding a Chern number of $C = 2$. This interplay among exchange splitting, SOC, and Berry curvature results in a sizable anomalous Hall conductivity of approximately 316~$\Omega^{-1}\mathrm{cm}^{-1}$. The system also exhibits a high Fermi velocity, $v_F \approx 0.78 \times 10^6$ m/s, indicative of efficient carrier transport. The estimated Curie temperature ($T_C \approx 3.5$ K) and in-plane magnetic anisotropy confirm broken time-reversal symmetry, which is essential for topological transport. Furthermore, anisotropies in the low-energy acoustic phonon spectrum exhibit directional correlations with Berry curvature features, suggesting that the vibrational dynamics are influenced by the underlying electronic topology. These results support 2D BaNiCl$_3$ as a candidate platform for dissipationless, spin-polarized transport arising from its intrinsic topological and magnetic features. Experimental verification via angle-resolved photoemission spectroscopy and magneto-optical Kerr effect measurements would provide further insight into the predicted topological and magnetic properties.

This research is supported by the U.S. Department of Energy, Office of Science, Basic Energy Sciences under Award DOE-SC0024099. This work used Stampede3 at the Texas Advanced Computing Center (TACC) through allocation PHY240252 from the Advanced Cyberinfrastructure Coordination Ecosystem: Services and Support (ACCESS) program, supported by NSF Grants 2138259, 2138286, 2138307, 2137603, and 2138296.



\end{document}


\LARGE
\textbf{Supplementary Materials: Intrinsic Chern Half-Metal with High Anomalous Hall Conductivity in 2D BaNiCl$_3$}\footnote{This research was supported by U.S. Department of Energy, Office of Science, Basic Energy Sciences under Award DOE-SC0024099 This work used Stampede3 at the Texas Advanced Computing Center (TACC) through allocation PHY240252 from the Advanced Cyberinfrastructure Coordination Ecosystem: Services and Support (ACCESS) program, supported by NSF Grants 2138259, 2138286, 2138307, 2137603, and 2138296.}\\[6pt]
\small
\textbf {Chidiebere I. Nwaogbo, and Chinedu E. Ekuma}\\[6pt]
Department of Physics, Lehigh University, Bethlehem, PA USA \\ cin221@lehigh.edu\\ che218@lehigh.edu\\[6pt]

\begin{figure}[htb!]
	\centering
	\includegraphics[trim = 0mm 0mm 0mm 0mm,width=0.5\linewidth,clip=true]{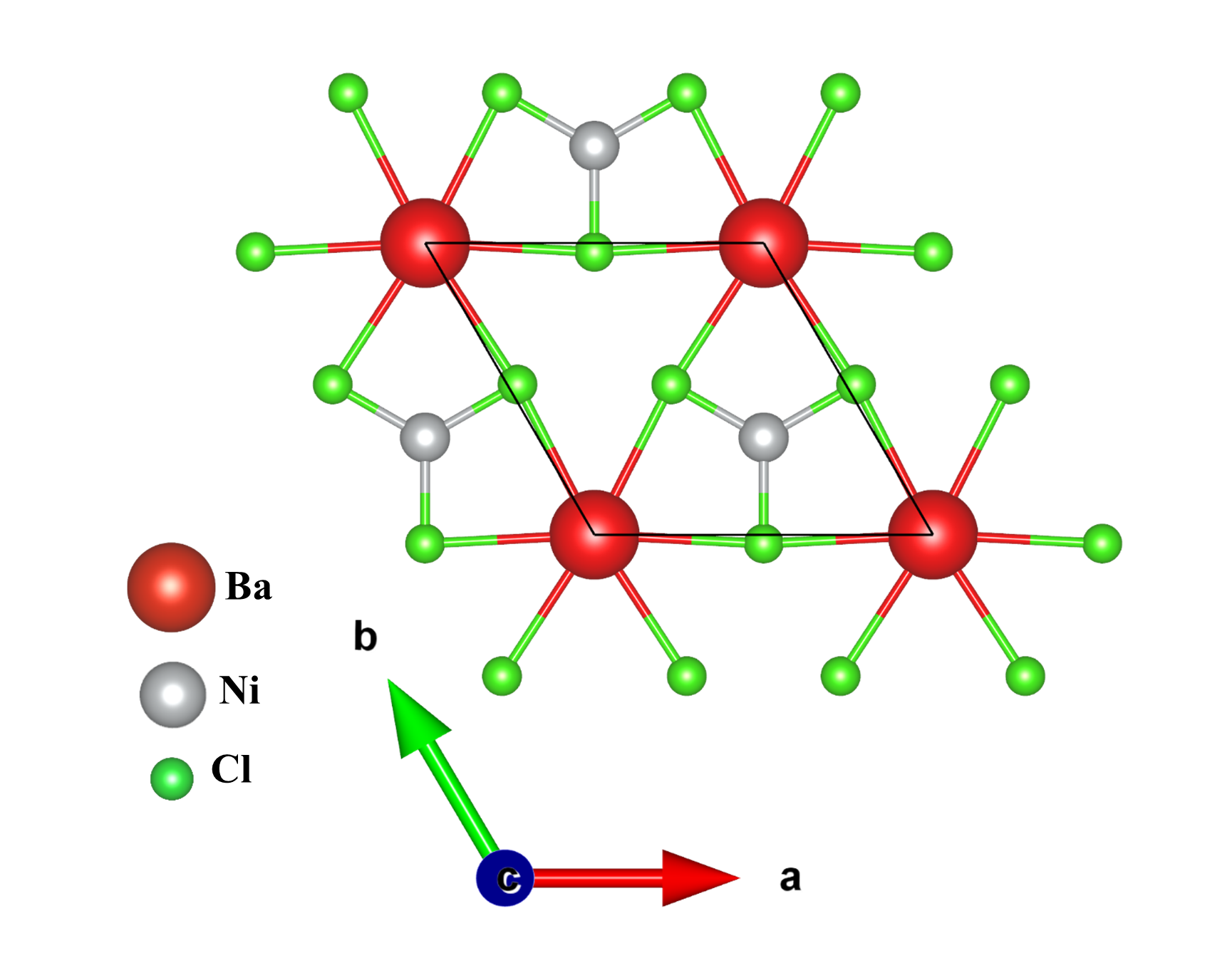} 
	\caption{Crystal structure of 2D-BaNiCl$_3$} 
	\label{figS1}
\end{figure}

We assess the structural stability of two-dimensional (2D) BaNiCl$_3$ by first performing geometry optimization. The relaxed structure reveals that Ba atoms preferentially occupy the corners of the lattice, while Ni atoms are positioned within Cl-centered octahedra, as illustrated in Figure~\ref{figS1}. The formation energy is calculated using the expression \(E_{\text{form}} = E_{\text{BNC}} - \sum n_i \epsilon_i,\) where $E_{\text{BNC}}$ denotes the total ground-state energy of the fully optimized 2D BaNiCl$_3$, $n_i$ represents the number of atoms of species $i$, and $\epsilon_i$ is the ground-state energy per atom of element $i$ in its standard bulk phase. The computed formation energy is approximately $E_{\text{form}} \approx -3.2~\text{eV}$, indicating an exothermic process and suggesting that the 2D structure is thermodynamically stable with respect to its elemental constituents.

\begin{figure}[htb!]
	\centering
	\includegraphics[trim = 0mm 0mm 0mm 0mm,width=0.5\linewidth,clip=true]{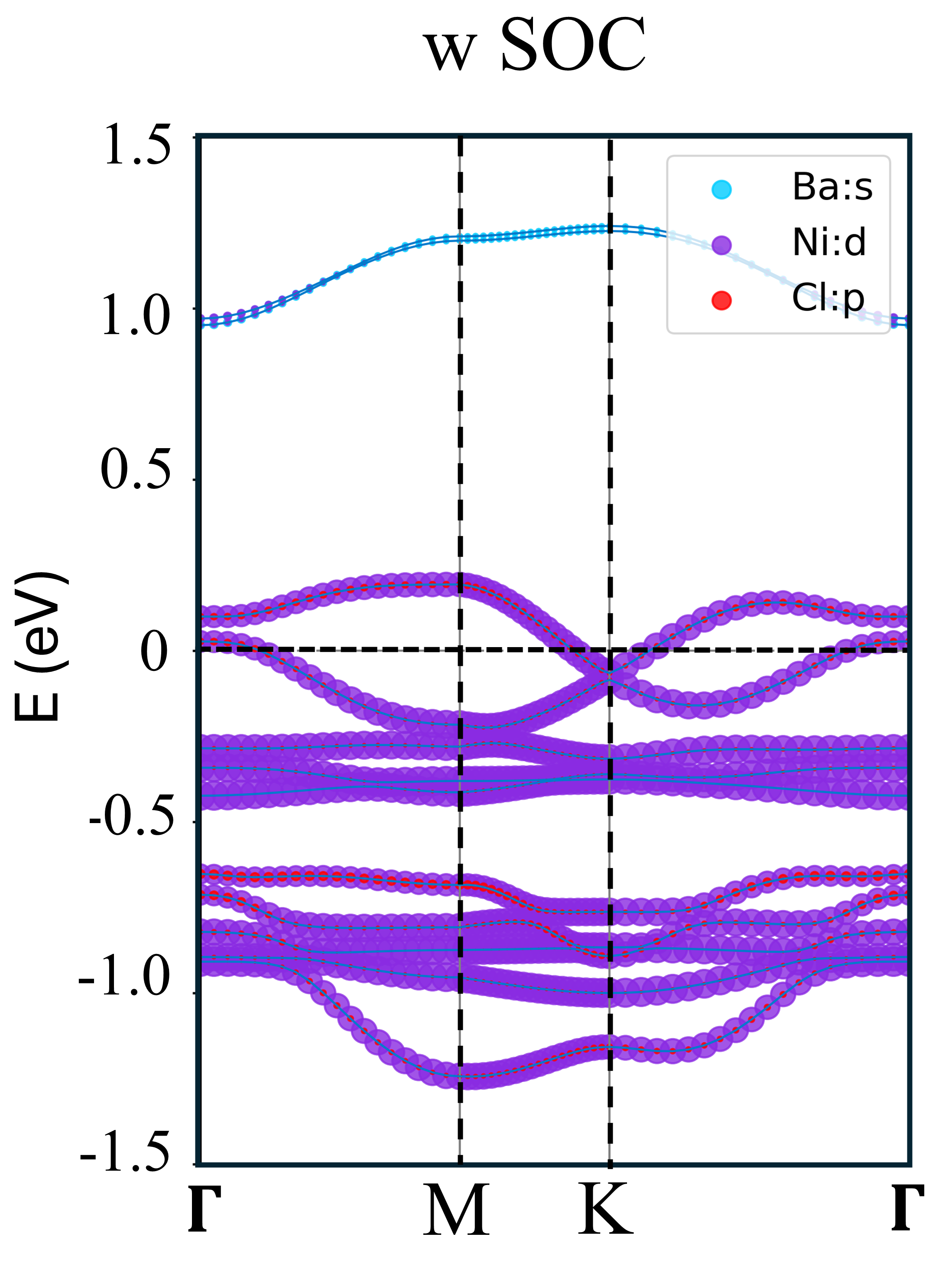} 
	\caption{ Projected band structrure of 2D BaNiCl$_3$ onto atomic orbitals. The blue, purple, and red points represent the contributions from Ba, Ni, and Cl atoms, respectively.} 
	\label{figS2}
\end{figure}

To further examine the feasibility of mechanical isolation, we compute the exfoliation energy using the method proposed by Jung \textit{et al.}~\cite{jung2018rigorous}, given by \(\Delta E_{x} = \frac{E_l(n) - E_{\text{bulk}}(n/m)}{A}\), where $E_l(n)$ is the total energy of the $n$-layer system, $E_{\text{bulk}}(n/m)$ is the energy per $n$ atomic layers extracted from the bulk structure (with $m$ total layers), and $A$ is the surface area of the bulk unit cell. The exfoliation energy of 2D BaNiCl$_3$ is found to be $\Delta E_{x} \approx 32.42~\text{meV}/\text{\AA}^2$, which lies within the range reported for easily exfoliable materials such as graphene, hexagonal boron nitride, MoS$_2$, and phosphorene (typically 18–32~meV/\AA$^2$), and is significantly lower than that of many other 2D materials evaluated via similar methods~\cite{barnowsky2023new,jung2018rigorous}. Moreover, this value falls within the upper bound for exfoliation energies proposed by Choudhary \textit{et al.}~\cite{choudhary2017high}, supporting the conclusion that 2D BaNiCl$_3$ is mechanically exfoliable and a viable candidate for experimental synthesis. Figure~\ref{figS2} presents the projected electronic band structure of 2D BaNiCl$_3$. The electronic states near the Fermi level are predominantly derived from Ni $d$ orbitals (shown in purple), whereas Cl $p$ orbitals contribute mainly to deeper valence bands.

While total energy calculations are commonly used to assess compound stability, they do not provide a complete picture of a material's overall stability. To address this limitation, we evaluated the mechanical properties of 2D BaNiCl$_3$ to determine its mechanical stability and assess its experimental viability. Specifically, we computed the two-dimensional (2D) elastic tensor elements and conducted a comprehensive analysis of the elastic and mechanical behavior using the Elastool toolkit~\cite{EKUMA2024109161,liu2022elastool}. The mechanical stability was assessed through the calculated elastic constants. The resulting values satisfy the Born stability criteria for hexagonal crystal systems~\cite{born1996dynamical,PhysRevB.90.224104}, thereby confirming that 2D BaNiCl$_3$ is mechanically stable.

\begin{table}[htb]
    \centering
    \begin{tabular}{|c|c|}
    \hline
     Model parameters & calculated values \\ \hline
    \textit{a,c} ($\AA$) & 6.75, 23.00 \\ \hline
    $E_{FORM}$(eV/atom)& -3.20 \\ \hline
    $\Delta E_x (eV/\AA^2)$& 0.0324\\ \hline
    $E_{FM}$(eV) &  -69.878 \\ \hline
    $E_{gAFM}$(eV) & -69.873 \\ \hline
    $E_{aAFM)}$(eV) & -69.874 \\ \hline
    $J_1$(meV/link) &0.365 \\ \hline
    $J_2$(meV/link)& 0.367\\ \hline
    \end{tabular}
    \caption{Computed Energy parameters for 2D BaNICl$_3$}
    \label{tab:S1}
\end{table}

The eigenvalues of the elastic tensor are \(0.217~\text{N/m}\), \(2.43~\text{N/m}\), and \(4.86~\text{N/m}\). The material exhibits a two-dimensional (2D) Young's modulus of \(0.42~\text{N/m}\) and a shear modulus of \(2.43~\text{N/m}\), indicating good resistance to both tensile and shear deformations. A Poisson's ratio of \(-0.91\) indicates auxetic behavior, meaning the material expands laterally when stretched and contracts laterally when compressed. The Pugh’s modulus ratio of \(0.04\) further classifies the material as brittle. Additionally, a strain energy density of \(0.00097~\text{J/m}^2\) reinforces the mechanical robustness of the material.

Dynamic stability is confirmed through phonon-related properties. The calculated sound velocities (\(V_l = 1.41~\text{km/s}\), \(V_s = 1.37~\text{km/s}\), and \(V_d = 1.39~\text{km/s}\)) and the Debye temperature (\(T_D = 134.61~\text{K}\)) indicate stable low-energy acoustic phonon behavior. Collectively, these mechanical and dynamical properties establish 2D BaNiCl$_3$ as a stable and promising candidate for advanced materials applications.

To estimate the Curie temperature (\(T_C\)), we first employed the mean-field approximation:
\[
T_C = \frac{2}{3} \cdot \frac{\Delta E}{z S N k_B} (S + 1),
\]
where \(\Delta E\) is the energy difference between the ferromagnetic (FM) and antiferromagnetic (AFM) states, \(S\) is the spin quantum number, \(z\) is the number of nearest neighbors, \(N\) is the number of magnetic atoms, and \(k_B\) is the Boltzmann constant. Following Hund’s rule, the spin quantum number is \(S = 1\) for the high-spin \(d^8\) configuration of Ni, where electrons fill the \(t_{2g}\) and \(e_g\) orbitals, leaving two unpaired electrons.

\begin{figure}[htb!]
	\centering
	\includegraphics[width=0.9\linewidth]{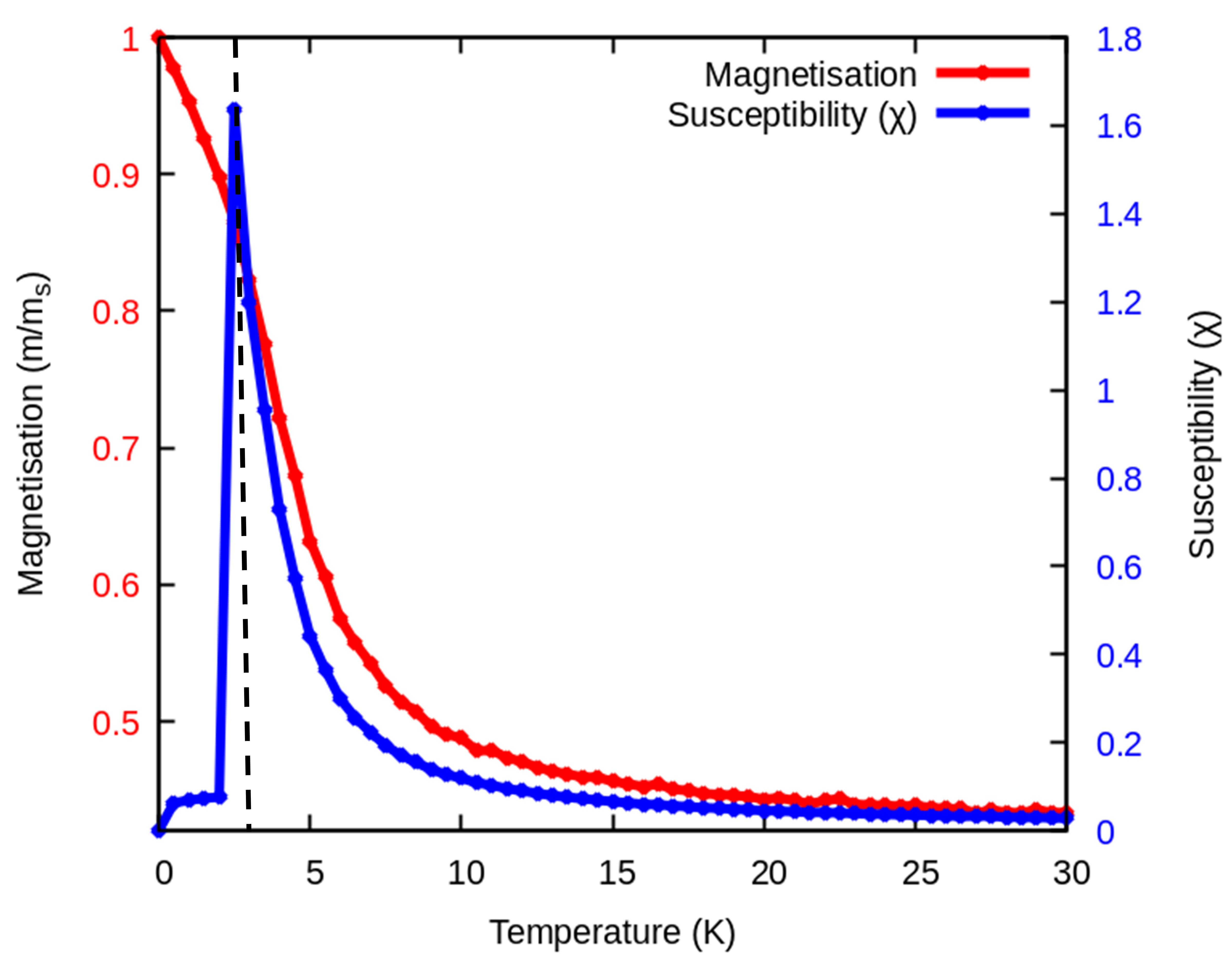} 
	\caption{Temperature dependence of (a) mean magnetization (red) and (b) magnetic susceptibility (blue) from Monte Carlo simulations.} 
	\label{figS3}
\end{figure}

To more accurately determine the magnetic transition temperature, we performed Monte Carlo simulations using the two-dimensional anisotropic Heisenberg model implemented in the VAMPIRE software package. The temperature dependence of mean magnetization and total magnetic susceptibility is shown in Figure~\ref{figS3}. The spin Hamiltonian used for this model is:
\begin{align}
H = -\frac{1}{2} J_1 \sum_{\langle i,j \rangle} \mathbf{S}_i \cdot \mathbf{S}_j 
    - \frac{1}{2} J_2 \sum_{\langle\langle i,j \rangle\rangle} \mathbf{S}_i \cdot \mathbf{S}_j 
    - K \sum_i |\mathbf{S}_i^z|^2,
\end{align}
where \(J_1\) and \(J_2\) are the exchange interaction strengths between nearest and next-nearest neighbors, respectively, \(K\) is the single-ion anisotropy constant, and \(\mathbf{S}_i\) is the spin vector at site \(i\).

The total energies of the FM, G-type AFM, and A-type AFM states are expressed as:
\begin{align}
E_{\mathrm{FM}} &= E_0 - 3J_1|\mathbf{S}|^2 - 3J_2|\mathbf{S}|^2 - A|\mathbf{S}|^2, \\
E_{\mathrm{gAFM}} &= E_0 + 3J_1|\mathbf{S}|^2 - 3J_2|\mathbf{S}|^2 - A|\mathbf{S}|^2, \\
E_{\mathrm{aAFM}} &= E_0 - A|\mathbf{S}|^2,
\end{align}
where \(E_0\) is the total energy of the non-magnetic state and \(A\) is the anisotropy contribution. The computed values used in these expressions are summarized in Table~\ref{tab:S1}.

The small magnitudes of \(J_1\) and \(J_2\) correlate with the low Curie temperature obtained. Although the relatively large magnetic anisotropy energy (MAE) could, in principle, stabilize long-range magnetic order through spin-wave gap opening, the low Curie temperature \(T_C = 3.5~\text{K}\) suggests that topology and spin-orbit-coupled fluctuations dominate over anisotropy-induced stability.

We computed the temperature-dependent mean magnetization using
\[
\langle |m| \rangle = \left| \frac{\sum_i \mu_i S_i}{\sum_i \mu_i} \right|,
\]
where \(\mu_i\) is the local spin moment, and the magnetic susceptibility as
\[
\chi_\alpha = \frac{\sum_i \mu_i}{k_B T} \left( \langle m_\alpha^2 \rangle - \langle m_\alpha \rangle^2 \right), \quad (\alpha = x, y, z),
\]
with \(m_\alpha\) denoting the directional components of magnetization. These quantities were sampled over 40,000 equilibrium time steps at each temperature. The temperature corresponding to the loss of magnetic alignment in the magnetization curve is identified as the Curie temperature.

